# Coherent Mode Decoupling: A Versatile Framework for High-Throughput Partially Coherent Light Transport

Han Xu[1], Ming Li[1], Shuo Wang[1], Zhe Ren[1], Peng Liu[1], Yi Zhang[1], Yuhui Dong[1] and Liang Zhou[1, 4]

[1]*Beijing Synchrotron Radiation Facility, Institute of High Energy Physics, Chinese Academy of Sciences, Beijing, People's Republic of China.*

[4]zhouliang@ihep.ac.cn.

# Abstract

Accurate and efficient wave-optics simulation of partially coherent light transport systems is critical for the design of advanced optical systems, ranging from computational lithography to diffraction-limited storage rings (DLSR). However, traditional approaches based on Coherent Mode Decomposition suffer from high computational costs due to the propagating massive sets of two-dimensional modes. In this paper, we propose the Coherent Mode Decoupling (CMDC) algorithm, a high-throughput computational framework designed to accelerate these simulations by orders of magnitude without compromising physical fidelity. The method factorizes 2D modes into efficient one-dimensional (1D) components, while crucially incorporating a subspace compression strategy to capture non-separable coupling effects. We demonstrated the generality and robustness of this framework in applications ranging from computational lithography to coherent beamlines of DLSR. For computational lithography, the convolution between kernels and masks is accelerated by factors ranging from 8 to 167 (maintaining $R^2$ of 99.9%–95.4%). Analysis of lens aberrations reveals that CMDC captures complex intensity distortions with a 30-fold speedup. For partially coherent beamline design, CMDC method achieved a speedup of approximately $10^3$ times, with the accuracy verified by ptychography experiments. This work provides a flexible and powerful tool for optical system design, tolerance analysis, and inverse design in computational lithography.

# 1. Introduction

Partially coherent light transport is essential for the performance of modern optical systems [1]. The impact of coherence is observed across various scales and wavelengths. In micro-scale applications such as Deep Ultraviolet (DUV) and Extreme Ultraviolet (EUV) lithography systems, partial coherence affects image resolution, requiring careful optimization of both the source and the mask [2-5]. Similarly, in optical microscopy, controlled coherence is key to achieving high contrast in phase-sensitive imaging [6-8]. On a larger scale, the development of Diffraction-Limited Storage Rings (DLSR) [9], such as the High Energy Photon Source (HEPS) [10], provides X-ray beams with high brightness and transverse coherence. In these facilities, accurate modeling of partial coherence is critical for designing beamlines that preserve the quality of the X-ray transport.

Given the complexity and cost of physical experiments, computational simulation has become an important tool for the design, optimization, and analysis of these optical systems. Computational lithography [3, 11, 12], for instance, relies on forward modeling of partially coherent imaging to predict resist patterns and apply Optical Proximity Correction. Similarly, for synchrotron beamlines, accurate wave-optics simulation is critical for designing long-distance beam transport systems that preserve the coherence properties of the source [13-15]. By virtually prototyping optical components, engineers can predict the impact of aberration and misalignments before fabrication [16, 17]. Consequently, developing efficient and accurate algorithms for partially coherent field propagation is not merely an academic pursuit but an industrial necessity to reduce development cycles and maximize the utility of systems across different optical fields.

However, partially coherent light calculation is computationally intensive [18-20]. The standard approach, Coherent Mode Decomposition (CMD), decomposed from the Cross-Spectral Density (CSD) function, representing the optical field as a sum of independent orthogonal modes [1]. To accurately describe the field, this method often requires decomposing the source into hundreds or even thousands of two-dimensional (2D) modes. Propagating each 2D mode individually through an optical system requires significant calculation time and memory. This high computational cost becomes a practical bottleneck for tasks that require many iterations, such as inverse design in computational lithography [18], tolerance analysis or iterative optimization of optical systems [20].

To address these challenges, we propose Coherent Mode Decoupling (CMDC) algorithm to accelerate partially coherent simulations. The core innovation of CMDC is its capability to decouple 2D coherent modes into separable 1D components, transforming the expensive 2D propagation problem into efficient 1D operations. Unlike simple approximations that assume perfect separability, the CMDC framework includes a robust residual correction mechanism. It extracts the non-separable "coupled energy" caused by optical imperfections or complex geometries and propagates this energy using a small set of 2D residual modes. This hybrid approach offers a flexible balance between speed and fidelity, making it suitable for rigorous simulations in lithography, optical microscopy, and DLSR beamlines, where both efficiency and precision are required.

## 2. Theoretical Framework

### 2.1 Cross-Spectral Density and Coherent Mode Decomposition

The statistical properties of a stationary, partially coherent optical field are fully characterized by CSD function $W(r_1, r_2, \omega)$, which describes the spatial correlation of the electric fields at two spatial points $r_1$ and $r_2$ at a frequency *w*. According to partial coherence theory [1], the CSD can be expanded into a sum of uncorrelated, orthogonal modes via CMD:

$$W(r_1, r_2, \omega) = \sum_n \lambda_n \phi_n(r_1, \omega) \phi_n^*(r_2, \omega) \quad (1)$$

where $\lambda_n$ are the eigenvalues representing the eigenvalue (energy) of each mode, and $\phi_n$ are the coherent eigenfunctions (modes). In computational wave-optics, propagating the CSD involves propagating each eigenmode $\phi_n(r, w)$ independently. For a 2D transverse field, $\phi_n(r, w)$ is a 2D complex array. However, the propagation of *N* such 2D coherent modes through an optical system is computationally expensive, with the complexity scaling as $O[N \times M_x M_y \log(M_x M_y)]$ when using propagators [21] utilizing Fourier transform (*FT*), where $M_x$ and $M_y$ are the pixel dimensions of 2D modes.

### 2.2. The Coherent Mode Decoupling (CMDC) Algorithm

To address the high computational cost of full 2D propagation, we introduce the CMDC algorithm. The fundamental assumption of this approach is to determine whether a complex 2D wave function $\phi_n(r)$ can be factorized into the product of two independent 1D functions, *i.e.*, $\phi_n(x,y) \approx \phi_n(x) \times \phi_n(y)$. In linear algebra, a discretized function satisfying this condition perfectly forms a matrix of rank 1. Therefore, the degree of "coupling" between the *x* and *y* coordinates is mathematically equivalent to the deviation of the mode matrix from a rank-1 approximation. We quantify and exploit this property by treating the discretized 2D coherent mode $\phi_n(x,y)$ as a complex 2D matrix and applying Singular Value Decomposition [22] (SVD):

$$\phi_n(x,y) = \sum_k \sigma_{n,k} u_{n,k}(x) v_{n,k}^*(y) = \sigma_{n,0} u_{n,0}(x) v_{n,0}^*(y) + \sum_{k=1} \sigma_{n,k} u_{n,k}(x) v_{n,k}^*(y) \quad (2)$$

Here, the SVD expands the 2D field into a weighted sum of orthogonal separable basis functions (outer products of vectors $u_{n,k}(x)$ and $v_{n,k}(y)$). The first term $u_{n,0}(x) v_{n,0}^*(y)$ corresponds to the largest singular value $\sigma_{n,0}$, represents the 1D decoupled modes $\phi_{n,sep}(x) = u_{n,0}(x)$ and $\phi_{n,sep}(y) = v_{n,0}(y)$, which are the optimal rank-1 approximation of the field in the least-squares sense [23]. This term captures the dominant, decoupled behavior of the wavefront. The remaining decomposed terms $\Delta\phi_{n,coupled}(x,y) = \sum_{k=1} \sigma_{n,k} u_{n,k}(x) v_{n,k}^*(y)$ constitute the coupled residual. These higher-order terms mathematically describe the features of the wavefront that cannot be represented by a simple Cartesian product, physically corresponding to the *x-y* coupling effects such as astigmatism, rotation, or diffraction from irregular apertures. Thus, the decomposition is defined as:

$$\phi_n(x,y) = \phi_{n,sep}(x) \times \phi_{n,sep}(y) + \Delta\phi_{n,coupled}(x,y) \quad (3)$$

By isolating the first term, we transform the majority of the 2D transport problem into two efficient 1D operations. The residual terms are not discarded but handled separately (as detailed in Section. 2.3), ensuring that the coupling information is preserved without carrying the full computational weight for the separable components.

## 2.3 Analysis and Residual Compression

The accuracy of the CMDC approximation is governed by the energy distribution between the separable and coupled parts. We define the Separable Energy Ratio for the $n^{th}$ mode as:

$$R_{sep,n} = \sigma_{n,0}^2 / \sum_k \sigma_{n,k}^2 \qquad (4)$$

Where $\sigma_{n,k}$ is the singular value obtained from Equation 2. For the entire partially coherent field, the total Global Separability Ratio is defined as the energy-weighted average of $R_{sep,n}$ over all modes:

$$R_{sep} = \sum_n \lambda_n R_{sep,n} / \sum_n \lambda_n \qquad (5)$$

where $\lambda_n$ is the eigenvalue of the $n^{th}$ coherent mode. For ideal Gaussian-Schell sources $R_{sep} = 1$. In practice, high $R_{sep}$ allows 1D propagation to capture the bulk of the partially coherent light transport. However, in the presence of optical imperfections or complex apertures, the residual term $\Delta\phi_{n,coupled}(x,y)$ cannot be ignored without loss of accuracy.

To maintain high fidelity when $R_{sep} < 1$, we implement a Subspace Compression strategy. Instead of propagating 2D residual terms $\Delta\phi_{n,coupled}(x,y)$ individually (which would be computationally intensive), CMDC method constructs a global residual matrix *C* by stacking the 2D residual terms:

$$C = [\sqrt{\lambda_0}\Delta\phi_0(x,y), \sqrt{\lambda_1}\Delta\phi_1(x,y), \sqrt{\lambda_2}\Delta\phi_2(x,y), \ldots \sqrt{\lambda_n}\Delta\phi_n(x,y)] \qquad (6)$$

Therefore, by performing a secondary decomposition on matrix *C*, the 2D "Coupled Modes" series:

$$C = [\beta_0\psi_0(x,y), \beta_1\psi_1(x,y), \beta_2\psi_2(x,y), \ldots \beta_m\psi_m(x,y)] \qquad (7)$$

are generated, where $\beta_m$ is the singular value. These modes capture the dominant features of the non-separable distortions. By propagating the top *M* 2D coupled modes $\psi_m(x,y)$ that satisfy a cumulative energy threshold (*e.g.*, 95%), we ensure the simulation maintains high fidelity with minimal additional computational cost. This hybrid approach ensures that the "non-separable" physics is preserved with minimal additional computation.

**2.4 Decoupling of Optical Elements**

The computational efficiency of CMDC requires the interaction between the optical field and the optical elements be performed in the one-dimensional domain. In the wave-optics framework, an optical element is usually characterized by a complex transmission function *T*(*x*, *y*), which modifies the amplitude and phase of the incident wavefront. Realistic optical systems often contain non-separable features. Examples include lithographic masks with complex circuit patterns (as demonstrated in the Section 3.1) or X-ray lenses with surface figure errors that break symmetry (Section 3.2). To address this, we apply the same decoupling strategy to the optical elements. The 2D transmission function $T(x, y)$ is decomposed into 1D components using SVD:

$$T(x, y) = \sum_k \gamma_k M_k(x) M_k^*(y) \tag{8}$$

where $\gamma_k$ are the singular values, and $M_k(x)$ and $M_k(y)$ are the decoupled 1D transmission profiles. By substituting this expansion into the propagation equation, the interaction of a single 1D decoupled mode $\phi(x)$ with a complex 2D element is transformed into a summation of 1D interactions:

$$\phi'(x) = \sum_k \sqrt{\gamma_k}\, \phi(x) M_k(x) \tag{9}$$

This formulation allows the CMDC algorithm to handle complex, non-separable optical elements while maintaining the speed advantage of 1D array operations.

## 3. Method Application and Verification

To evaluate the practical performance and versatility of the proposed CMDC framework, we applied the algorithm to three different optical systems ranging from micro-fabrication to large-scale X-ray facilities. Firstly, a computational lithography simulation was conducted to benchmark the computational efficiency of the method against traditional 2D approaches. Secondly, we analyzed an optical system incorporating measured surface height error of a Compound Refractive Lens (CRL) [24] to test the robustness of the algorithm against the realistic wavefront distortions. Thirdly, to verify the physical accuracy of the method, we conducted a full wave-optics simulation of the Hard X-ray Coherent Scattering (HXCS) [15] beamline at the High Energy Photon Source (HEPS) and compared the results with experimental data measured at the facility. The following sections detail the simulation setup and results for each case.

### 3.1. Computational Efficiency in Lithography

To benchmark the computational efficiency of the CMDC algorithm in handling high-dimensional optical fields, we simulated a standard lithography configuration (Figure 1a). The simulation parameters were set to a wavelength of $\lambda$ = 193 nm and a numerical aperture (NA) of 0.3, representing a typical configuration for EUV lithography system [18]. The illumination source was defined as an annular pupil with a coherence factor of 0.69 (Figure 1b), which requires a high number of coherent modes to accurately represent the partial coherence. The object was a binary mask containing complex circuit patterns (Figure 1c), discretized onto a computational grid of 1024×1024 pixels.

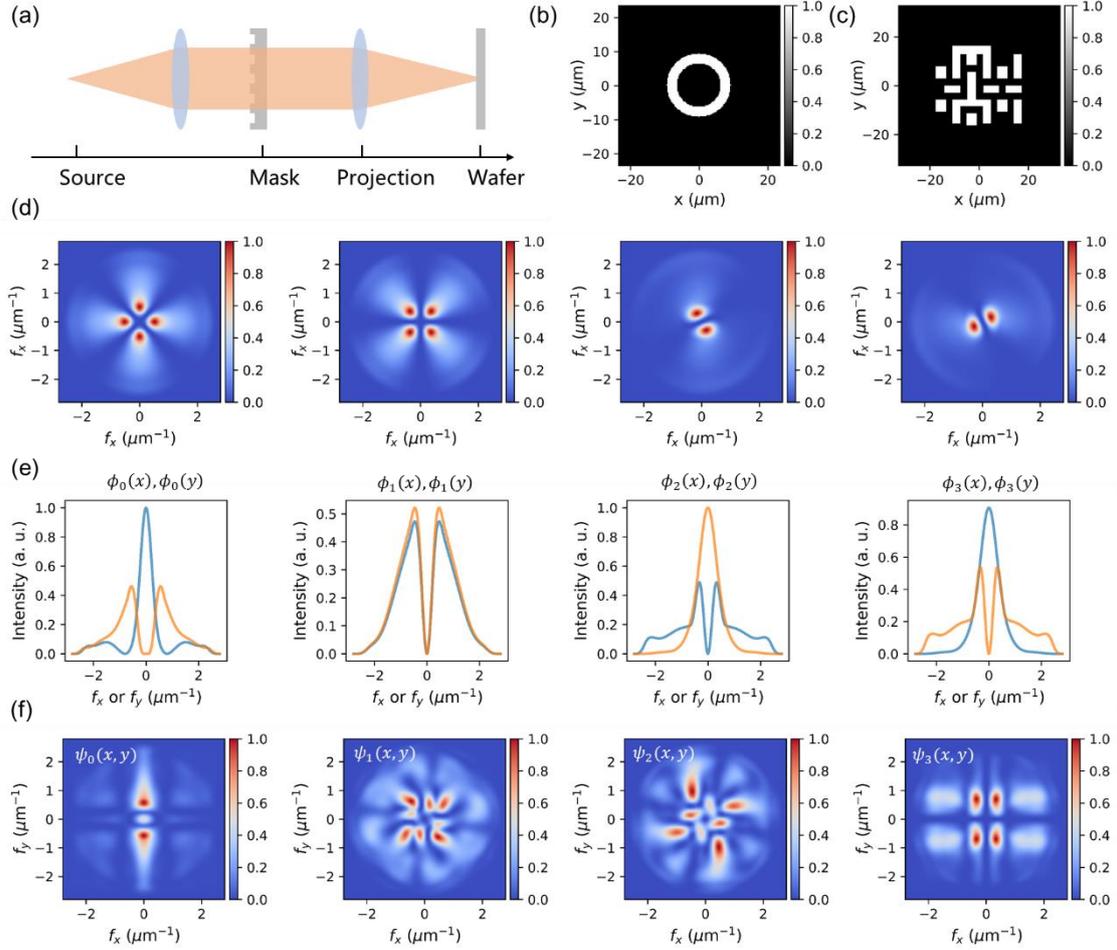

Figure 1. (a) The optics layout for the lithography system. (b) The annular source. (c) The binary pattern of the mask. (d) The first four decomposed 2D kernels $K_n(x, y)$. (e) The decomposed decoupled 1D kernels $\phi_n(x)$ (blue) and $\phi_n(y)$ (orange). (f) The 2D coupled modes $\psi_n(x, y)$.

The simulation of lithography system employed the Sum of Coherent Systems (SOCS) method [3, 11], where the Transmission Cross Coefficient (TCC) was first decomposed into a series of orthogonal eigenfunctions (kernels) $K_n(x, y)$. Figure 1d displays the first four dominant 2D kernels derived from the TCC, and 61 kernels are employed with an occupation larger than 98.3%.

Using the CMDC framework, we further decomposed these 2D kernels into 1D decoupled modes $\phi_n(x)$ and $\phi_n(y)$ using the decoupling strategy described in Section 2.2, and the intensity profiles are shown in Figure 1e, which exhibit symmetric profiles, capturing the primary optical characteristics of the illumination. The $R_{sep}$= 0.58, which suggests that 1D decoupled modes propagation was able to maintain 58% total energy of the source. The remaining 42% energy of the source is represented by the 2D coupled modes (Figure 1f), which could be used to adjust the accuracy of the CMDC method.

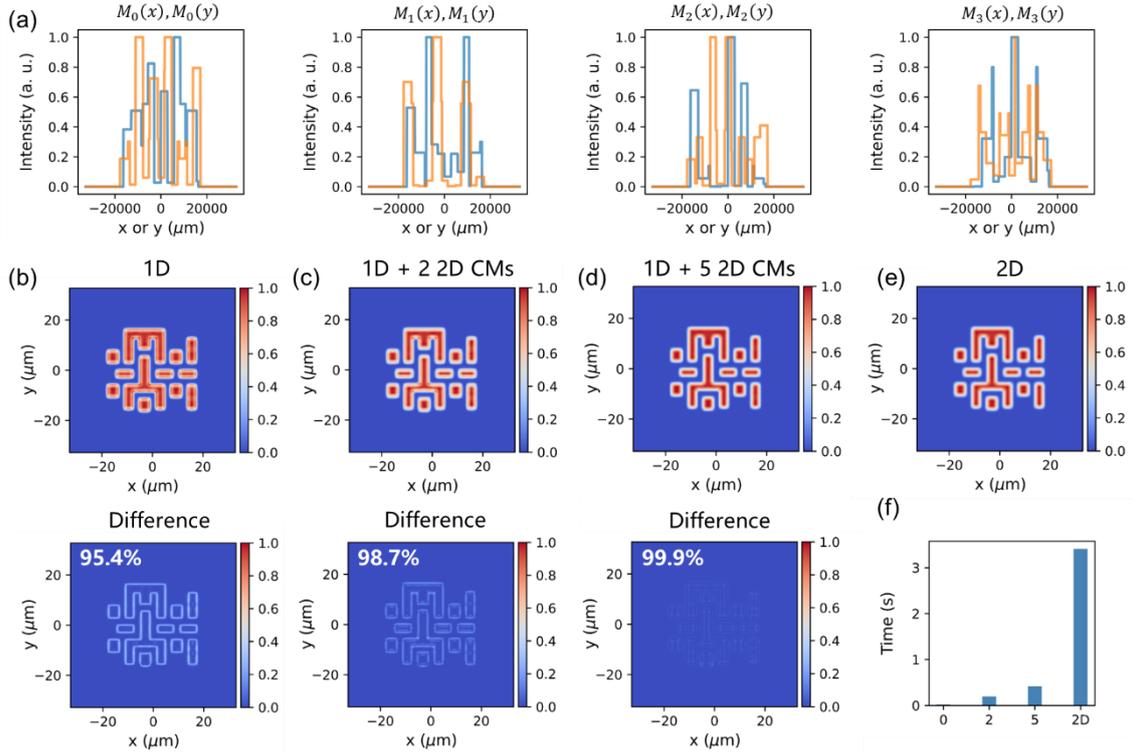

Figure 2. (a) The decomposed decoupled 1D mask modes $M_n(x)$ (blue) and $M_n(y)$ (orange). The aerial images calculated by the convolution using 1D decoupled kernels and (b) 0, (c) 2 and (d) 5 2D coupled kernel modes. CM: coupled mode. The difference is also shown and the $R^2$ is labeled. (e) The aerial images calculated by SOCS method (2D kernels) (f) The computational times for the convolution with CMDC and SOCS method. The labels "0", "2" and "5" represents the computational time of 1D decoupled kernels and 0, 2 and 5 2D coupled kernels. The label "2D" represents the computational time of SOCS method.

We then calculated the aerial image intensity using both the traditional 2D SOCS method and the proposed CMDC method (1D decoupled modes and 2D coupled modes). For 1D decoupled modes propagation, the mask is decomposed into 1D components using the method described in Section 2.4. The decomposed singular values reveal that the first six 1D modes $M_k(x)$ and $M_k(y)$ (Figure 2a) already have an occupation of 99%. We compared the result with 2D kernel modes using 1D decoupled modes. As shown in Figure 2b, the visual features of the circuit patterns, including the corners and dense lines, are reproduced identically by both methods. To quantify the accuracy, we computed the difference map between the two aerial images, and the $R^2$ = 95.4%. Though the 1D decoupled kernel modes already produce aerial images with $R^2$ greater than 95%, we test the effect of 2D coupled kernel modes to adjust the accuracy, considering only 58% energy is presented by 1D decoupled modes. Thus, 2 and 5 2D coupled kernel modes are employed to further correct the aerial images, and the energy occupation of source has increased to 93.4% and 99.7%. As shown in Figure 2c and d, the difference between CMDC method and 2D kernel modes is decreased to $R^2$ = 98.7% and 99.9%.

The computational performance is compared in Figure 2f. The calculation time for the full 2D convolution between kernels and mask was 3.4 seconds. In contrast, the convolution between kernels and mask using 1D CMDC method complete in 0.02 seconds. With 2 and 5 more 2D coupled kernels, the computational time is increased to 0.20 and 0.41 s. This corresponds to a speedup factor of approximately 167 ($R^2$ = 95.4%), 17 ($R^2$ = 98.7%) and 8 ($R^2$ = 99.9%). This result indicates that for computational lithography, the CMDC framework allows high-speed, accuracy adjustable calculation for computational lithography without compromising image fidelity, which benefits high-throughput calculation such as ILT.

### 3.2. Evaluation of Lens Aberrations

Optical aberrations are a fundamental factor limiting the performance of focusing systems, reducing the peak intensity and degrading the image quality [25]. In the design and analysis of partially coherent optical systems, evaluating the impact of surface errors is essential for tolerance analysis. While analytical methods can estimate metrics like the Strehl ratio, rigorous computational wave-optics simulation is required to accurately predict the detailed intensity distribution, such as the asymmetric side-lobes induced by realistic manufacturing defects.

In this section, we employed the CMDC framework to evaluate the focusing performance of a refractive lens under partially coherent illumination. The light source was modeled as a Gaussian-Schell Modes (GSM), a standard model for describing partially coherent beams with Gaussian intensity and coherence profiles (Figure 3a and Figure S1). To reflect realistic experimental conditions, we incorporated measured surface height error data from a fabricated refractive lens (Figure 3b, has been transform to phase error). The measured error profile was scaled to simulate a strongly aberrated lens with a Strehl Ratio of 0.5. Though GSM is completely separable along *x-y* axes, a 2D surface shape error introduces a complex phase modulation term $e^{i\phi(x,y)}$, which typically breaks the Cartesian symmetry of the optical system and 2D GSMs propagation is required. Based on CMDC method, we apply the decoupling strategy described in Section 2.4 directly to this 2D phase error. As shown in Figure 3c, the 2D phase error is decomposed into a series of 1D decoupled modes $M_n(x)$ and $M_n(y)$, and the first 15 phase error modes already have an occupation of 99%.

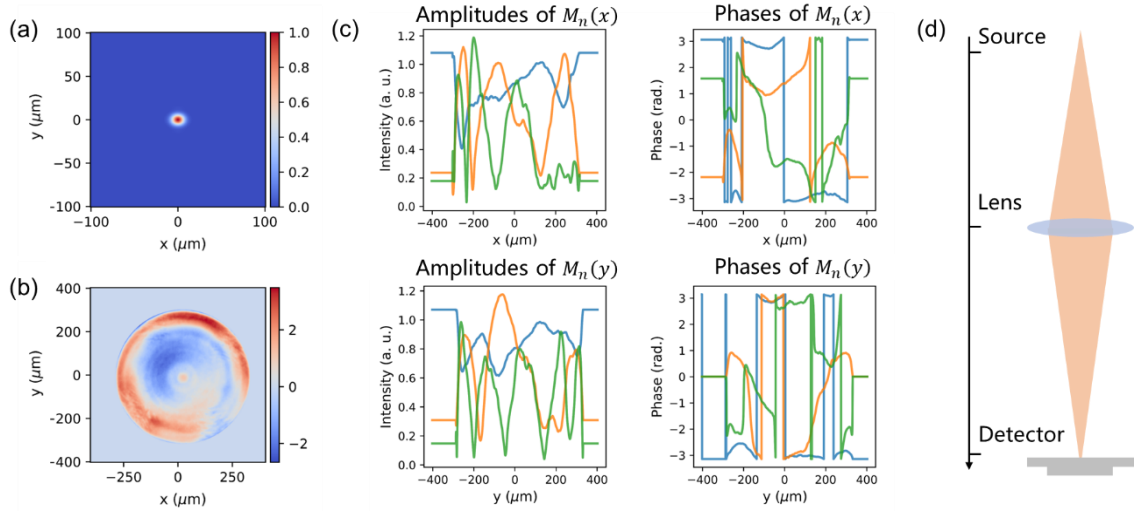

Figure 3. (a) The intensity of the partially coherent GSM light source (see Supplementary, Figure S1 for the details). (b) The phase error $\phi(x,y)$ of the lens. $\phi(x,y) = 2\pi\delta h(x,y)/\lambda$, where $\lambda$ is wavelength, $\delta$ is refractive index, $h(x,y)$ is the measured surface height error. (c) The amplitudes and phases of the decomposed 1D phase error modes $M_n(x)$ and $M_n(y)$. First three modes (1-3, blue, orange, green) are revealed. (d) A schematic diagram of a direct focusing optical system for aberration analysis.

A direct focusing optical system (Figure 3d) was employed to verify both the accuracy and efficiency of the CMDC method. The results are presented in Figure 4. As shown in Figure 4a, the focus calculated using only the zero-order GSM (representing the fully coherent limit) reveals intensity distortions induced by the significant phase error. Figure 4b displays the result obtained using a superposition of 100 GSMs, while the partial coherence smooths the strong distortions, the characteristic asymmetry remains evident. Crucially, the results calculated by the proposed 1D CMDC method shows strong agreement with those obtained from the 2D propagation for both the coherent and partially coherent cases, which is consistent with the 99% occupation for the first 15 phase error modes $M_n(x)$ and $M_n(y)$. The cross-sectional intensity profiles further confirm that the fast 1D decoupled calculation achieves high fidelity compared to the full-wave simulation. In terms of computational cost, the traditional 2D method required 153.6 s, whereas the CMDC method completed the calculation in only 4.7 s. This confirms that the CMDC algorithm effectively captures the physics of lens aberrations, allowing for efficient tolerance analysis of non-ideal optical elements in general focusing systems.

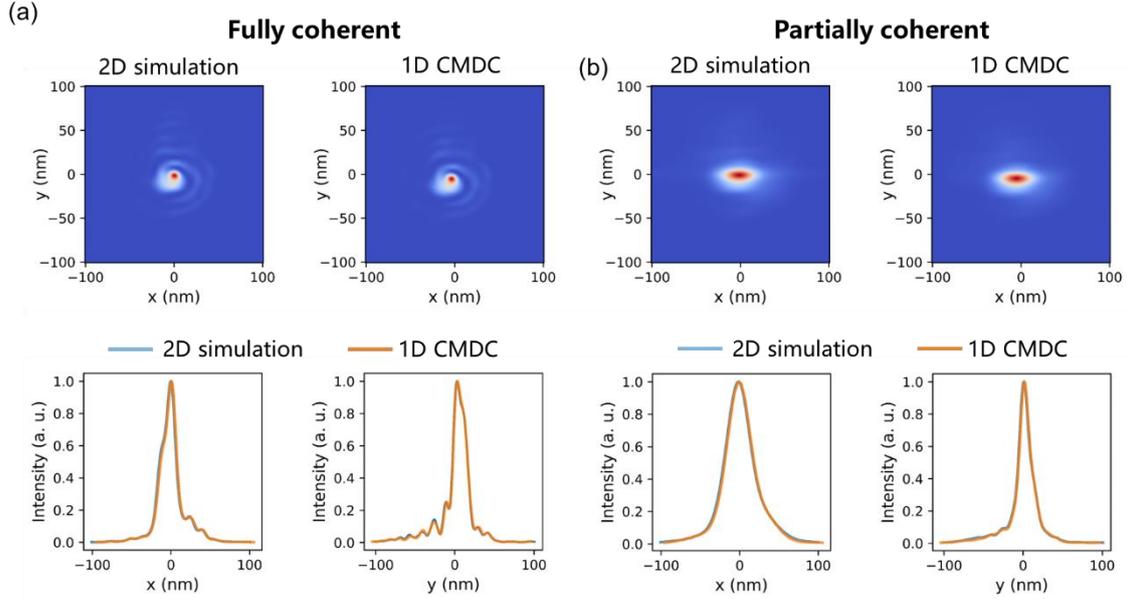

Figure 4. Comparison of focal intensity distributions calculated using traditional 2D propagation and the proposed 1D CMDC method. (a) Fully coherent using zero-order GSM, where the 2D intensity maps and corresponding cross-sectional profiles show significant distortions induced by phase errors. (b) Partially coherent using 100 GSMs, demonstrating how partial coherence partially smooths the distortions while maintaining characteristic asymmetry. The cross-section profiles are shown below.

## 3.3. Experimental Validation using Ptychography Experiment.

To verify the physical accuracy of the CMDC framework within the context of general partially coherent optics, we conducted experimental validation at the Hard X-ray Coherent Scattering (HXCS) beamline [15] at the High Energy Photon Source (HEPS) as an example. This X-ray optical system, in a Diffraction-Limited Storage Ring (DLSR), represents a distinct class of partially coherent light transport systems [14]. The short wavelength (angstrom level) challenges the accuracy of wave-optics simulation. The undulator insertion generates a partially coherent wavefield that is fundamentally described by statistical optics theory, making this coherent beamline an ideal platform for benchmarking high-precision wave-optics algorithms of simulation.

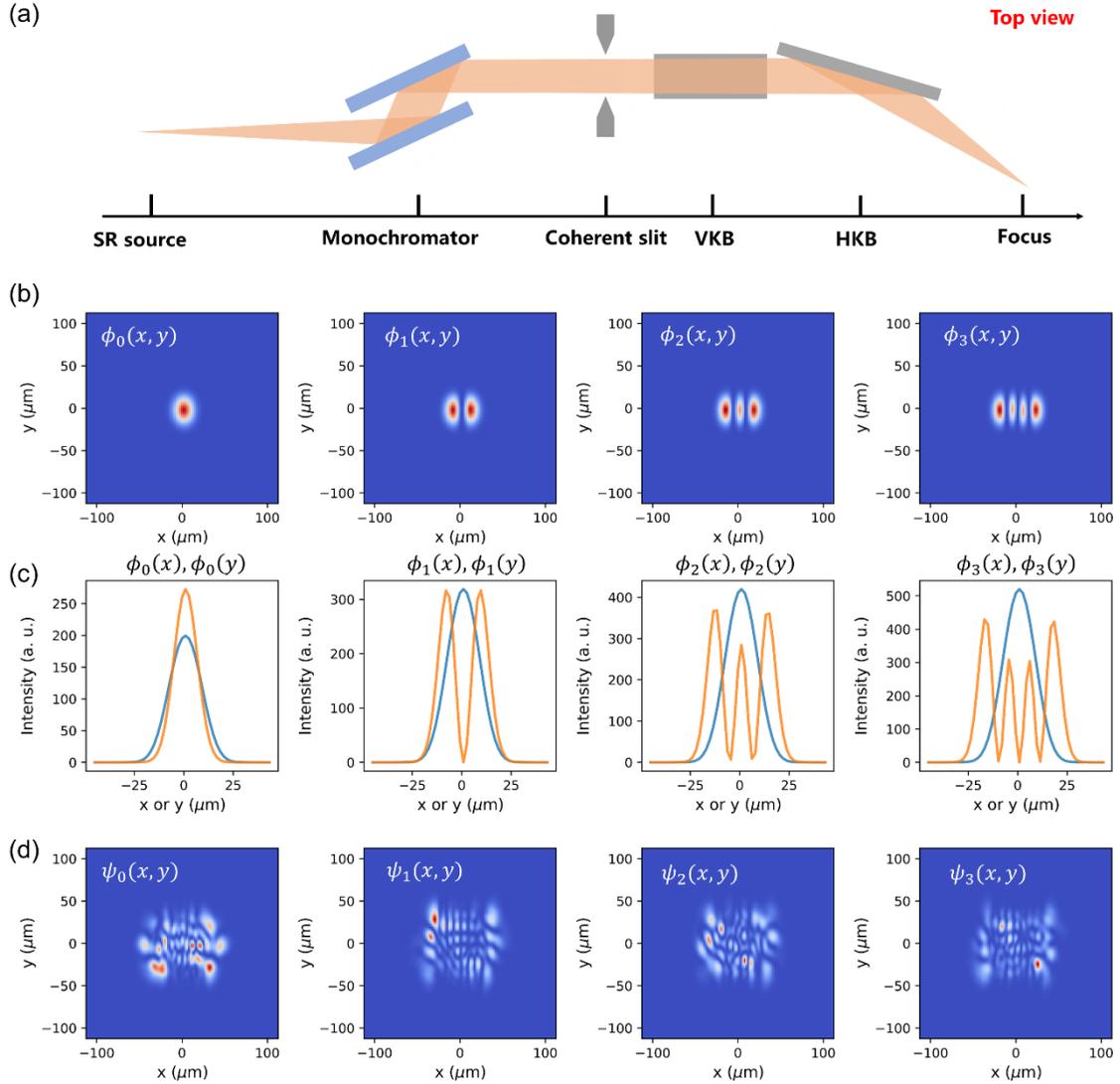

Figure 5. (a) A schematic diagram of an optics layout for HXCS beamline (Top view). SR source: Synchrotron Radiation source; VKB: Vertical KB mirror; HKB: Horizontal KB mirror. (b) The first four 2D coherent modes $\phi_n(x,y)$ of SR source. (c) The 1D decoupled modes $\phi_{n,sep}(x)$ (blue) and $\phi_{n,sep}(y)$ (orange). (d) The first four 2D coupled modes $\psi_m(x,y)$.

The optical layout (Figure 5a) consists of the undulator source [15] operating at 12.4 keV, a Double Crystal Monochromator (DCM, 40 m) for energy resolution (temporal coherence). A coherence-defining slit for spatial coherence manipulation (79.1 m) with an opening of 152 $\mu$m (horizontal) × 425 $\mu$m (vertical), defines a numerical aperture of 84.4 $\mu$rad (horizontal) × 236 $\mu$rad (vertical). A pair of Kirkpatrick-Baez (KB) mirrors [26] for nanofocusing are placed at 79.62m and 79.84 m. This optical configuration challenges computational optics algorithms as it involves a long-distance propagation (80 m), diffraction by a rectangular aperture, and high-NA nanofocusing by reflective optics. The high-NA nanofocusing required dense sampling for the accuracy when using wave-optics propagators like Fresnel propagator, which results in high computational cost [27].

The coherent modes $\phi_n(x, y)$ for SR source are shown in Figure 5b, which exhibit a partially coherent light source with a coherent fraction of 20.7%. After the application of CMDC method, the intensities of the 1D decoupled modes $\phi_{n,sep}(x)$ and $\phi_{n,sep}(y)$ are generated (Figure 5c). As expected, the horizontal and vertical profiles differ significantly but are smooth and well-behaved, indicating that the physical asymmetry of the source aligns with the Cartesian basis. Consequently, the 2D coupled modes $\psi_m(x, y)$ are shown in Figure 5d. The quantitative analysis confirms a high $R_{sep}$ = 0.95, suggesting that 1D decoupled modes propagation could achieve high accuracy [28].

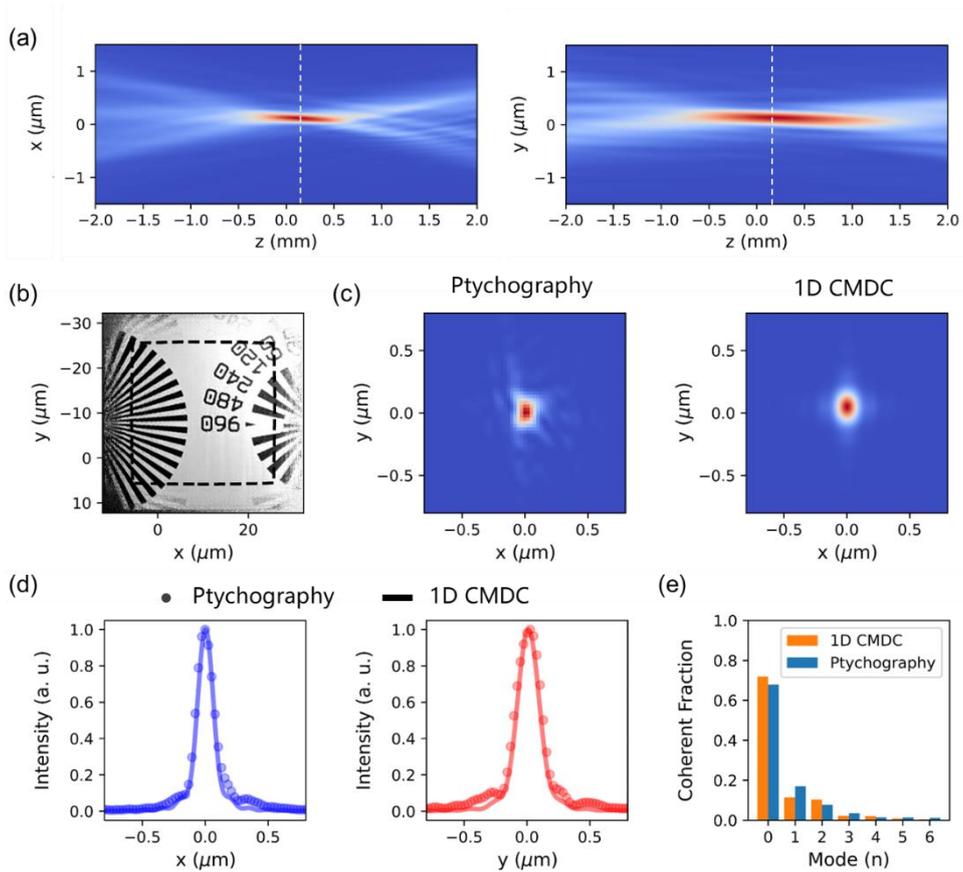

Figure 6. (a) The back-propagation of wavefront from sample to the focus. (b) The Siemens stars test target. (c) The 2D intensity distribution of focus from ptychography experiment (left) and simulation (right). (d) The 1D intensity distribution (integral) of focus from ptychography experiment (spot) and simulation (solid

line). (e) The coherent fraction of the different modes of focus from ptychography (blue) and simulation (orange).

The 1D decoupled modes (Figure 5c) are propagated through the optics to the focus. Totally 3.4 s was consumed for the propagation, which reveal high efficiency compared with the propagation of 2D coherent modes (68 min). To rigorously validate the simulation, we compared the results with experimental measurements obtained from the beamline. The wavefront at the sample is reconstructed using ptychography, and back-propagated to the focus [29] (Figure 6a). And Siemens stars (Figure 6b) test target is employed. The measured intensity distribution shows consistency with the CMDC simulation results (FWHM horizontal 144 nm × vertical 198 nm), as shown in Figure 6c and d. The difference might be attributed to beamline imperfections, such as surface errors of the mirrors and environmental vibrations. Furthermore, the coherence of the beam was quantitatively verified. As shown in Figure 6e, the coherent fractions of the modes derived from the multi-modal ptychographic reconstruction [30] (first 7 modes) shows excellent agreement with the prediction from simulation by CMDC method. These results, which match in both intensity and coherence, confirm that the CMDC algorithm accurately captures the physics of partially coherent light transport.

## 4. Conclusion

In this work, we have introduced the CMDC algorithm to meet the requirement for high-throughput, high-fidelity wave-optics simulation in partially coherent light transport systems such as computational lithography, optical microscopy and DLSR X-ray beamlines. By decomposing the 2D propagation problem into a series of 1D operations, CMDC addresses the fundamental computational bottleneck of traditional CMD methods. The practical utility of our framework is exploiting the dominant separability of most optical fields (1D decoupled modes) while preserving essential coupling effects via a subspace compression strategy (2D coupled modes). Our benchmarks in computational lithography demonstrate that CMDC can accelerate the kernel-mask convolution process in aerial image calculations by a factor of 167 with a high accuracy ($R^2$ = 95.4%). Furthermore, the analysis of lens aberrations reveals that CMDC effectively captures the intensity distortions induced by realistic surface errors, achieving a 30-fold speedup compared to 2D wave-optics analysis. Finally, the experimental validation at the HEPS HXCS beamline confirms that the algorithm accurately captures the complex modal structure of X-ray sources, reducing simulation times from over an hour to 3.4 s. Thus, CMDC offers a versatile and scalable tool for optical engineers. By providing a tunable balance between speed and precision, it opens new possibilities for real-time tolerance analysis, large-scale inverse design, and the optimization of next-generation partially coherent imaging systems.

## Acknowledgments.

This work was supported by High Energy Photon Source (HEPS), a major national science and technology infrastructure in China.

## Disclosures.

The authors declare no conflicts of interests.

## Data availability.

Data underlying the results presented in this paper are not publicly available at this time but may be obtained from the authors upon reasonable request.

See Supplementary for supporting content.